\documentclass[11pt]{article}

\usepackage{fullpage}
\usepackage{setspace}
\usepackage{parskip}
\usepackage{titlesec}
\usepackage[section]{placeins}
\usepackage{xcolor}
\usepackage{breakcites}
\usepackage{lineno}
\usepackage{hyphenat}

\usepackage{caption}
\usepackage{subcaption}
\usepackage{multirow}
\newcommand*\rot{\rotatebox{90}}

\PassOptionsToPackage{hyphens}{url}
\usepackage[colorlinks = true,
            linkcolor = blue,
            urlcolor  = blue,
            citecolor = blue,
            anchorcolor = blue]{hyperref}
\usepackage{etoolbox}

\usepackage[round]{natbib}
\let\cite\citep

\renewenvironment{abstract}
  {{\bfseries\noindent{\abstractname}\par\nobreak}\footnotesize}
  {\bigskip}

\titlespacing{\section}{0pt}{*3}{*1}
\titlespacing{\subsection}{0pt}{*2}{*0.5}
\titlespacing{\subsubsection}{0pt}{*1.5}{0pt}

\usepackage{authblk}

\usepackage{graphicx}
\usepackage[space]{grffile}
\usepackage{latexsym}
\usepackage{textcomp}
\usepackage{longtable}
\usepackage{tabulary}
\usepackage{booktabs,array,multirow}
\usepackage{amsfonts,amsmath,amssymb}
\providecommand\citet{\cite}
\providecommand\citep{\cite}

\newif\iflatexml\latexmlfalse

\AtBeginDocument{\DeclareGraphicsExtensions{.pdf,.PDF,.eps,.EPS,.png,.PNG,.tif,.TIF,.jpg,.JPG,.jpeg,.JPEG}}

\usepackage[utf8]{inputenc}
\usepackage[english]{babel}

\usepackage{float}

\date{}

\begin{document}

\title{Asset Allocation: From Markowitz to Deep Reinforcement Learning}

\author[1]{Ricard Durall}%

\affil[1]{Open University of Catalonia}

\maketitle

\begin{abstract}
Asset allocation is an investment strategy that aims to balance risk and reward by constantly redistributing the portfolio's assets according to certain goals, risk tolerance, and investment horizon.
Unfortunately, there is no simple formula that can find the right allocation for every individual.
As a result, investors may use different asset allocations' strategy to try to fulfil their financial objectives.
In this work, we conduct an extensive benchmark study to determine the efficacy and reliability of a number of optimization techniques.
In particular, we focus on traditional approaches based on Modern Portfolio Theory, and on machine-learning approaches based on deep reinforcement learning.
We assess the model's performance under different market tendency, i.e., both bullish and bearish markets.
For reproducibility, we provide the code implementation code in this \href{https://github.com/RicardDurall/Benchmarking-Strategies-for-Asset-Allocation}{repository}.
\end{abstract}

\noindent\textbf{Keywords:} Asset Allocation, Portfolio Selection, Markowitz Portfolio, Deep Reinforcement Learning, Benchmarking\\

\section{Introduction}
Asset allocation is an investment strategy that aims to balance risk and reward by apportioning a portfolio's assets according to an individual's financial situation, risk tolerance, investment horizon, and goals.
Fixed asset allocation refers to the portfolio that remains the same until the investor, or the portfolio's manager on behalf of the investor, decides to change the portfolio.
The ``60/40 portfolio'' is a well-known fixed allocation strategy that has been employed as a trusty guidepost for moderate risk investors.
It essentially consists of allocating 60\% of the portfolio to equities, and the other 40\% to bonds and other fixed-income instruments.
Another popular fixed asset allocation is the equal weight method.
This strategy gives the same importance to each asset in the portfolio, independent of external events.
Although fixed allocation approaches have been revered for their simplicity and reliability, the field of asset allocation still has a lot of room for improvement.
The mean-variance optimization model, proposed by Markowitz \cite{markowitz1968portfolio} serves as the keystone to Modern Portfolio Theory (MPT).
It is a practical method for selecting investments to maximize their overall returns within an acceptable level of risk; the Sharpe ratio \cite{sharpe1998sharpe} is one of its most employed add-ons to measure the risk-adjusted return with respect to the risk-free asset.
However, Markowitz model applies historical return and volatility as a proxy for future expectations in the allocation model.
Consequently, the model can make inaccurate assumptions since the returns and variances will likely not be the same in the future \cite{merton1980estimating}.
\cite{black1992global} solved this issue by applying an equilibrium return, based partially on Capital Asset Pricing Model (CAPM) \cite{sharpe1964capital, lintner1975valuation}, as a baseline for defining the expected return vector.
Further alternatives are the equal volatility portfolio, that uses the same amount of volatility in every asset; minimum variance portfolio, \cite{haugen1991efficient, chopra2013effect}, that provides the lowest variance among all possible portfolios of risky assets; maximum diversification portfolio \cite{choueifaty2008toward, choueifaty2013properties}, that maximizes the ratio of weighted-average asset volatilities to portfolio volatility; maximum decorrelation portfolio \cite{christoffersen2012potential}, that maximizes the diversification ratio based on the correlation matrix; and risk parity \cite{maillard2010properties, roncalli2016risk}, that uses the concept of the Security Market Line (SML) as part of its approach; being SML a graphical representation of the CAPM.
Besides the debatable assumption about the stationarity in time of the market, another drawback from Markowitz's approach appears when the amount of different assets is large enough.
MPT uses the covariance to determine which assets are included in the portfolio.
This statistical measure has quadratic growth with the number of asset and thus, those portfolios with rich diversity of assets will inevitably suffer from computational problems.
To circumvent this issue, \cite{bai2009enhancement} proposed to employ the theory of the large-dimensional random matrix.
Finally, another reason for poor performance of the mean-variance portfolio might be caused by the symmetry of asset returns.
\cite{low2016enhancing} showed that it is possible to enhance Markowitz's portfolio selection by allowing distributional asymmetries.

With the increasing use of artificial intelligence, deep-learning approaches have achieved remarkable breakthroughs leading to state-of-the-art results in various domains such as computer vision \cite{krizhevsky2012imagenet, goodfellow2014generative}, natural language processing \cite{vaswani2017attention, devlin2018bert}, and speech recognition \cite{deng2013new,chiu2018state}.
Such remarkable success has also sparked the interest of the finance research community.
As a result, in the past years, the number of deep-learning applications for portfolio asset allocation has dramatically increased.
\cite{lin2006recurrent} designed a dynamic portfolio selection model by incorporating the Recurrent Neural Network (RNN) \cite{rumelhart1985learning}.
\cite{freitas2009prediction, niaki2013forecasting, nguyen2015sentiment, heaton2017deep} relied on deep neural networks to model the market's behaviour so that the final solution found the optimal asset allocation.
A similar approach by \cite{chakravorty2018deep} dealt with macroeconomic data in conjunction with price-volume data in a walk-forward setting.
\cite{obeidat2018adaptive} proposed to predict future portfolio's returns using Long Short-Term Memory (LSTM) \cite{schmidhuber1997long} neural networks, as they are a more suitable than RNNs for processing time-series (sequential data).
Nonetheless, all the previous models lack interaction with the market.
In other words, they cannot adapt and consequently, they underperform in non-stationary scenarios.
To address this limitation, reinforcement-learning-based systems are suitable candidates. \cite{almahdi2017adaptive} introduced a recurrent-reinforcement-learning method, with a coherent risk-adjusted performance objective function, named the Calmar ratio, to obtain both buy and sell signals that updated the asset allocation weights.
\cite{jiang2017cryptocurrency} presented a Convolutional Neural Network (CNN) \cite{lecun1998gradient} to dynamically optimize cryptocurrency portfolios.
A follow-up work by \cite{jiang2017deep, i2020deep} assessed the impact of different types of layers, including CNN, LSTM and RNN.
Similarly, \cite{liang2018adversarial} applied deep-reinforcement-learning algorithms with continuous action space to asset allocation.
In particular, they investigated the model-free Deep Deterministic Policy Gradient (DDPG) \cite{silver2014deterministic} as well as the Proximal Policy Optimization (PPO) \cite{schulman2017proximal}.
\cite{buehler2019deep} presented a framework to hedge a portfolio of derivatives, where the system did not depend on specific market dynamics, such as transaction costs, market impact, liquidity constraints or risk limits.
\cite{kolm2020modern} investigated the link between portfolio allocation and reinforcement learning, namely, showing how the latter could be used to solve intertemporal financial problems.
\cite{yang2020deep} introduced an ensemble strategy that employed various deep-reinforcement schemes to learn a unified strategy that maximized the investment return.
\cite{ye2020reinforcement} proposed a framework, coined state augmented reinforcement learning, that aimed to leverage additional diverse information from alternative sources other than classical structured financial data like asset prices.
Finally, driven by the Covid-19 financial crisis, \cite{benhamou2021detecting} focused on models that detected extreme negative patterns, and consequently dis-investing the assets.

\section{Methodology}
First, we present the traditional approaches, where we discuss different strategies based on the Markowitz model.
Then, we describe deep reinforcement learning and its paradigm for asset allocation.

\subsection{Markowitz Mean-Variance Portfolio Theory}
Modern Portfolio Theory \cite{markowitz1968portfolio} introduces a financial method, for risk-averse investors, to construct diversified portfolios that optimize their returns.
Namely, the mean-variance optimization approach.
This technique aims at assembling a portfolio given some pre-defined constraints.
Risk and return trade-off is at the heart of such a method, and its main components are the standard deviation and expected return of the assets.
While variance (derived from the standard deviation) expresses the degree of spread in the data set, by showing how spread out the returns of a specific asset are, the expected return represents the probability of the estimated return of the investments.
MPT makes several key assumptions that the practitioners should be aware of before using the mean-variance optimization. The main ones are the following:

\begin{itemize}
    \item The risk of the portfolio is based on its volatility of returns, i.e., price fluctuations.
    \item The analysis is conducted on a single-period model of investment.
    \item Investors are rational, averse to risk and eager to increase consumption. As a result, the utility function is concave and increasing. 
    \item Investors seek either to maximize their portfolio return for a given level of risk or to minimize their risk for a given return.
\end{itemize}

From a mathematical perspective, given a portfolio $p$ with $n$ assets, we can calculate the standard deviation as:
\begin{equation}
\sigma_p = \sqrt{\sigma_p^2} ,
\end{equation}

the variance as:
\begin{equation}
\sigma_p^2 = \sum_{i=1}^n \sum_{j=1}^n \omega_i \omega_i Conv(r_i,r_j) = \omega^\mathrm{T} \Omega \omega,
\end{equation}

and the expected return as:
\begin{equation}
\mathbb{E}(r_p) = \sum_{i=1}^n \omega_i \mathbb{E}(r_i).
\end{equation}

The variable $\omega$ denotes the weights of the individual assets, and the variable $r_p$ the return of the portfolio.

\begin{figure}[t!]
  \includegraphics[width=\linewidth]{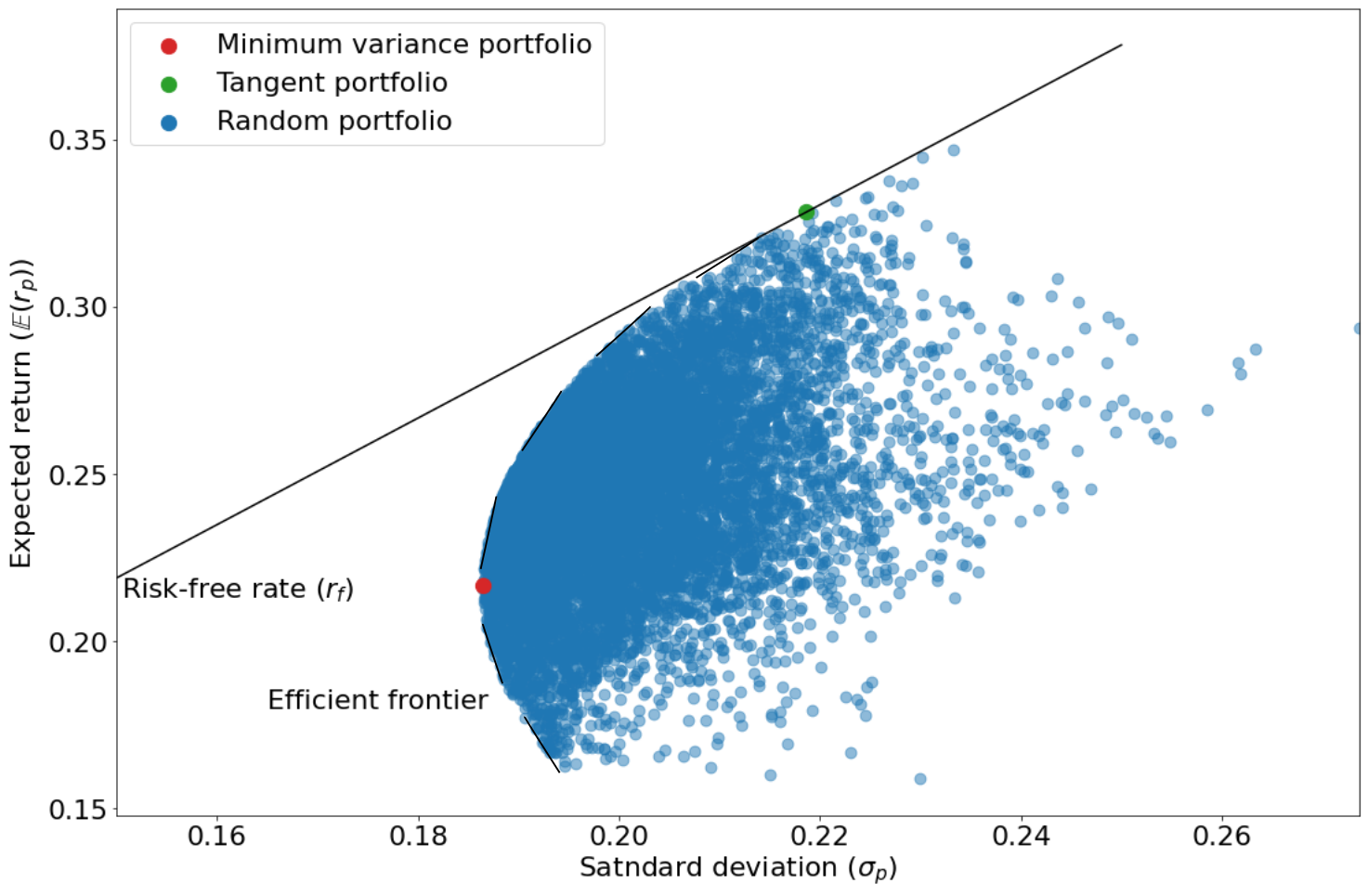}
  \caption{This graph shows the different types of portfolios based on Markowitz model.
  Each portfolio is plotted according to its expected return and standard deviation.
  If a portfolio is plotted on the right side of the chart, it indicates that the level of risk is too high for its return.
  If it is plotted low on the graph, the portfolio offers low returns for the risk that it carries.}
  \label{fig:markowitz}
\end{figure}

As we mentioned before, the trade-off between the risk and the expected return will be critical when building a portfolio.
Using the Markowitz model for analysing portfolios helps to discover the efficient frontier, which is the combination of assets that offers the highest expected return for a defined level of risk, or the lowest risk for a given level of expected return (see Figure \ref{fig:markowitz}).
Portfolios that lie below the efficient frontier are suboptimal, as they do not provide enough return for the level of risk.
Portfolios that cluster to the right of the efficient frontier are also suboptimal because they have a higher level of risk for the defined rate of return.
In this work, we are interested in two important points along the efficient frontier: minimum variance portfolio and tangency portfolio.

\subsubsection{The Minimum Risk Mean-Variance Portfolio}
The point where the hyperbola (efficient frontier) changes from convex to concave is where the minimum variance portfolio lies. This portfolio allocation has a unique solution that can be found by solving a simple quadratic optimization problem via standard Lagrange multiplier methods. The optimization problem can be formulated as:
\begin{equation}
\begin{aligned}
    & \min_{\omega}
    & & \frac{1}{2} \omega^\mathrm{T} \Omega \omega \\
    & \text{subject to}
    & & \omega^\mathrm{T} r = \mathbb{E}(r_p) \\
    & \text{and} 
    & & \omega^\mathrm{T} 1 = 1.
\end{aligned}
\end{equation}

The vector $\omega$ denotes the individual investments (weights of the assets) subject to the condition that the available capital is fully invested, i.e., $\mathrm{\omega}^\mathrm{T} 1 = 1$.
The lower bound on the target return $\mathbb{E}(r_p)$ is expressed by the condition $\mathrm{\omega}^\mathrm{T} r = \mathbb{E}(r_p)$, where the vector r estimates the expected mean of the assets ($\mathbb{E}(r_i)$).

\subsubsection{The Tangency Portfolio}
The tangency portfolio is the asset allocation that maximizes the Sharpe ratio \cite{sharpe1998sharpe}. This ratio measures the excess return earned over the risk-free rate per unit of volatility or total risk, which helps investors to better understand the return of their investment. It can be formulated as:
\begin{equation}
\begin{aligned}
    & \mathrm{Sharpe \; ratio} = \frac{\mathbb{E}(r_p) - r_f}{\sigma_p}
\end{aligned}
\end{equation}

where $r_f$ stands for risk-free rate, i.e., the theoretical rate of return of an investment with zero risk like U.S. treasury rate. 
The tangency portfolio optimization can be formulated as:
\begin{equation}
\begin{aligned}
    & \max_{\omega}
    & & \frac{\omega^\mathrm{T} r- r_f}{\omega^\mathrm{T}
    \Omega \mathrm{\omega}} \\
    & \text{subject to}
    & & \omega^\mathrm{T} 1 = 1.
\end{aligned}
\end{equation}

Graphically, it is the point where a straight line through the $r_f$ is tangent to the efficient frontier, in the Markowitz model space. 

\subsubsection{The Risk Parity Portfolio}
Risk parity \cite{maillard2010properties,roncalli2016risk} is an alternative approach to the Markowitz model that focuses on the allocation of the risk instead of the capital. This method asserts that when asset allocations are adjusted to the same risk level, the portfolio can achieve a higher Sharpe ratio and thus, it can be more resistant to market downturns. To achieve that, the risk parity portfolio tries to constrain each asset to contribute equally to the portfolio overall volatility. The optimization problem can be formulated as:
\begin{equation}
\begin{aligned}
    & \min_{\omega}
    & & \frac{1}{2} \omega^\mathrm{T} \Omega \omega - \frac{1}{n} \mathrm{ln}(\omega) \\
    & \text{subject to}
    & & \omega^\mathrm{T} 1 = 1.
\end{aligned}
\end{equation}

\subsection{Deep Reinforcement Learning}
Machine learning is a branch of artificial intelligence that allows machines to learn from data, identify patterns and make decisions without being explicitly programmed for it. 
Reinforcement learning is an area of machine learning that focuses on training an algorithm following the cut-and-try approach.
More specifically, the algorithm needs to learn to take actions that maximize the final reward in a particular situation.
To that end, this algorithm (agent) evaluates a current situation (state), takes an action, and receives feedback (reward) from the environment.
Positive feedback is given when the action is correct, and negative feedback otherwise.
Similar to other mathematical methods, machine-learning algorithms have different flavours, each of them with its own advantages and disadvantages.
Commonly, these algorithms can be divided into supervised, unsupervised and reinforcement learning. Supervised learning works with labelled data and mainly deals with regression and classification tasks.
The unsupervised learning, however, employs unlabelled data and tries to determine patterns and associations within the data.
This technique tackles clustering and associative rule mining problems.
Finally, reinforcement learning uses a learning agent to interact with the environment based on an action-reward system through a trade-off between exploitation and exploration.
The main goal of this type of learning is to find the best sequence of decisions that maximizes the long-term reward.
Due to the absence of training data, reinforcement-learning algorithms are bound to learn from their experience.
In particular, they learn how to act best through many attempts and failures.

\subsubsection{Actor Critic Optimization}
Actor-Critic (AC) \cite{konda1999actor} is a temporal difference method that has two separated memory structures to explicitly represent the policy and the value function.
On the one hand, the policy structure is known as the actor because it decides which action should be taken.
It essentially controls how the agent behaves by learning the optimal policy $\pi$ (policy-based).
On the other hand, the estimated value function is known as the critic.
It evaluates the action made by the actor by computing the value function (value-based), which can be the action-value $q_\pi$ or state-value $v_\pi$.
These two structures participate in an optimization game, where they both get better in their own role as the time passes.
The outcome is that the overall method (system) achieves superior results than systems based on solely one structure.

The Advantage Actor-Critic (A2C) \cite{mnih2016asynchronous} algorithm is a variation of AC, where the algorithm specifically uses estimates of the advantage function for its bootstrapping, i.e., to update a value based on some estimates and not on some exact values.
The function of the advantage function is to determine how good an action compared to average action for a specific state is.
By doing so, the variance between the old and the new policies is reduced, and consequently the stability of the reinforcement-learning algorithm improves.
The advantage function can be defined as:
\begin{equation}
a_\pi(s,a) = q_\pi(s,a) - v_\pi(s).
\end{equation}

Deep Deterministic Policy Gradient (DDPG)\cite{lillicrap2015continuous} is another AC method. It combines ideas from Deterministic Policy Gradient (DPG) \cite{silver2014deterministic} and Deep Q-Network (DQN) \cite{mnih2013playing}.
Namely, DDPG uses a critic that learns from a temporal loss and an actor that learns using policy gradient. However, DDPG is an off-policy method.
This means that it can sample batches from large experience buffers, making this approach more sample-efficient, at least at training.

Although DDPG can provide excellent results, it is frequently brittle with respect to hyperparameters and other kinds of tiresome fine-tuning dependencies.
Furthermore, a common failure of DDPG is that this algorithm continuously overestimates the $q_\pi$ values of the critic network, and it can eventually lead to the agent falling into a local optimum or to a catastrophic forgetting.
Twin Delayed DDPG (TD3) \cite{fujimoto2018addressing} is an algorithm that addresses this issue by introducing three novel critical tricks: (1) employing two critic networks, (2) delaying the updates of the actor and (3) adding noise to regularize the target action.

Soft Actor-Critic (SAC) \cite{haarnoja2018soft} is an algorithm that optimizes a stochastic policy in an off-policy fashion, forming a bridge between stochastic policy optimization and DDPG-based approaches.
The biggest feature of SAC is its modified objective function, where instead of only seeking to maximize the lifetime rewards, the algorithm also tries to maximize the entropy of the policy.
We can think of entropy as how unpredictable a random variable is. For instance, if a random variable always takes a single value, then it has zero entropy.
As a rule of thumb, we want a high entropy in our policy to explicitly encourage exploration, by assigning equal probabilities to actions that have same or nearly equal $q_\pi$ values, and to ensure that it does not collapse.

Another alternative to modify policy gradient methods using experiences from old versions of the policy is through the importance sampling technique.
This technique weights samples based on the difference between the action probability distributions, given by the current and old policies.
Trust region policy optimization (TRPO) \cite{schulman2015trust} is one policy gradient method that guarantees that the new update's policy is not far away from the old policy, or at least, that the new policy is within the trust region of the old policy.
However, in practice, TRPO is a relatively complicated algorithm to implement and not always a suitable candidate.
Proximal Policy Optimization (PPO) \cite{schulman2017proximal} is a follow-up work that simplifies the algorithm.
More specifically, PPO is a first-order optimization that defines the probability ratio between the new and old policies.
Instead of adding complex constraints, e.g., Kullback-Leibler, PPO imposes a policy ratio to stay within a small interval around 1. It can be viewed as a combination of A2C (having multiple workers) and TRPO (using trust region to improve the actor).

\subsubsection{Deep Reinforcement Learning for Asset Allocation}
Portfolio management can be modelled as a deep-reinforcement-learning (DRL) problem, where deep neural networks are used to optimize the portfolio risk-adjusted returns for a given set of assets.
To guarantee convergence, these neural networks need to understand the market's behaviour when reallocating the assets.
The DRL framework for portfolio management can be built as follows:

\begin{figure}[h!]
  \centering
  \includegraphics[width=.65\linewidth]{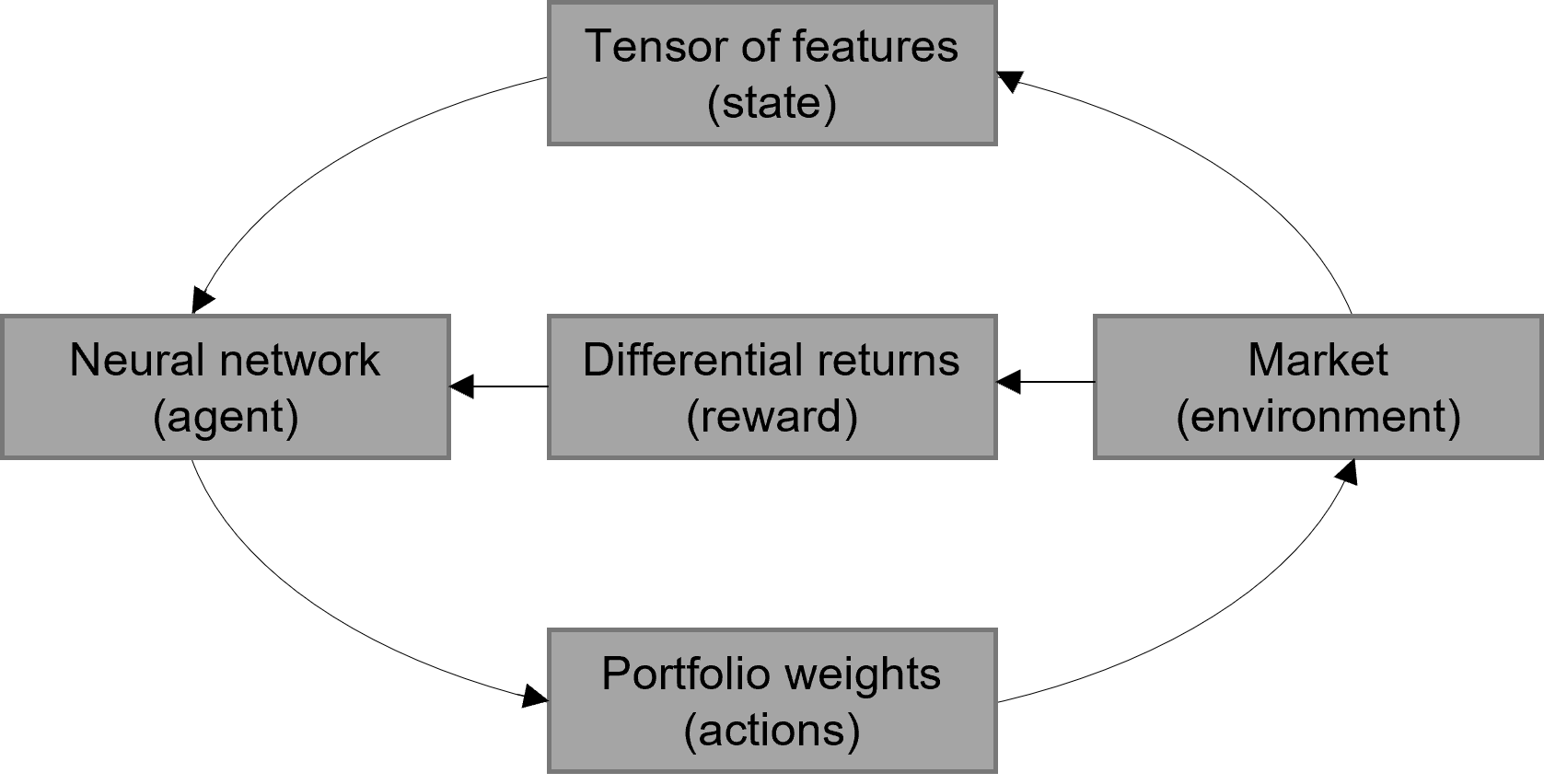}
  \caption{Representation of our deep-reinforcement-learning framework.}
  \label{fig:rdl}
\end{figure}

\emph{Agent}: A deep neural network is proposed as the agent, whose goal is to find an optimal function (either policy- or value-based) that learns actions that maximize the reward.

\emph{Actions}: They are the outputs of the network for a given period, and they represent the final portfolio weights, i.e., the percentage of each asset within the portfolio.

\emph{State}: It is the input used to feed the network and describes the current situation of the stock market via financial indicators.
These indicators are made of a tensor of features, which includes the current balance of our portfolio, the stock prices, and the owned assets.

\emph{Environment}: The stock market acts as the environment. 
It receives the actions taken by the agent and sends back the reward to the agent.

\emph{Reward}: The difference between the previous and the current portfolio values is our reward (differential returns).

In a DRL approach, usually the agent uses the stochastic gradient descent, based on a market snapshot (state) and on the reward (cost-adjusted returns), to learn the action (weights) that leads to the optimal portfolio allocation for a given point in time.
In Figure \ref{fig:rdl}, we can see the connection among the different elements of a DRL framework.
Finally, we need to make a few assumptions for a correct interpretability of our results:
\begin{itemize}
    \item It is possible to trade at the market any time.
    \item Our transactions do not to affect the market price of the assets. 
\end{itemize}

\section{Experiments}
In this work, we address the task of asset allocation. Given a portfolio with a set of $n$ assets, we aim to benchmark several allocation optimization methods, including traditional and DRL ones.
To that end, we conduct a comprehensive evaluation, where we investigate the robustness of such approaches for different market's conditions as well as different time frequencies of reallocation.
In particular, we study bull and bear market scenarios, on a day-trading setup.

\subsection{Experimental Settings}
When the exchanges close, the last trading price of the stock is recorded as the closing price of the share.
Nonetheless, it is not always reliable since it might not provide an accurate picture of the true value.
Therefore, the closing price is adjusted considering factors such as dividends, stock splits, and new stock issues, originating the adjusted close price.
In all our experiments, we use adjusted close price as input.

For each market scenario, we select a set of 8 stocks, from well-known companies, that follow the market tendency that we target to study.
We use data from a period of seven years, where 80\% of the dataset goes into the training set, and the remaining 20\% of the dataset goes into the testing set.
Notice that for the traditional methods, where no learning is involved, we do not have such a split.
Furthermore, for these methods, we set a windows size of 50 and no transaction costs are involved.
On the contrary, for the DRL-based counterpart, we use a windows size of 1 and their costs are set to 0.1\% for each trade total value.

We evaluate the performance of 9 different algorithms: tangency portfolio, minimum variance portfolio, risk parity, equal weight, A2C, PPO, DDPG, SAC and TD3.
While the first four approaches are deterministic, the last five approaches, based on the Actor-Critic algorithm, are not due to the (stochastic) weight initialization process.
Thus, we report results from 10 independent runs to obtain uncertainty boundaries. 

\clearpage
\subsection{Bull Market Scenario}
Bull market occurs when investment prices rise for a sustained period of time.
Propelled by the thriving economies and low unemployment rate, investors are eager to buy or hold onto securities.
The result is a buyer's market.
Bull markets tend to last for months or even years. Famous bull markets were the 1970s economic recovery as well as the pre-global financial crisis bull market.

\begin{figure}[h]
  \centering
  \includegraphics[width=\linewidth]{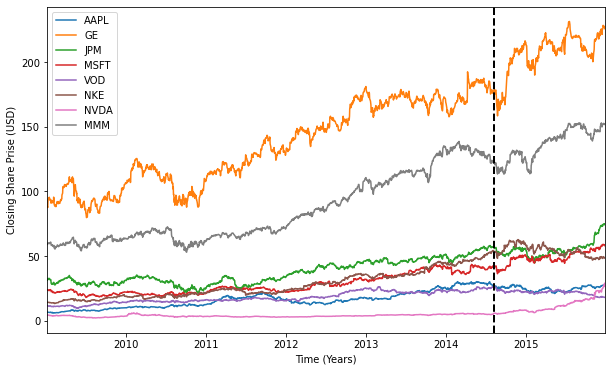}
  \caption{Evolution of the closing share price of our set of assets during bull market; from the 1st of January 2010 to the 1st of January 2017.
  The vertical dashed line separates the training data (left) from the testing data (right).}
  \label{fig:bull_data}
\end{figure}

Our first case of study focuses on 8 stocks that follow a bullish trend. 
These securities belong to the following companies:  Apple (APP), General Electric (GE), JPMorgan Chase (JPM), Microsoft (MSFT), Vodafone Group (VOD), Nike (NKE), Nvidia (NVDA) and 3M (MMM).
We use adjusted close prices from the 1st of January 2010 to the 1st of January 2017 (see Figure \ref{fig:bull_data}).

\begin{figure}
     \centering
     \begin{subfigure}[b]{\textwidth}
         \centering
         \includegraphics[width=\textwidth]{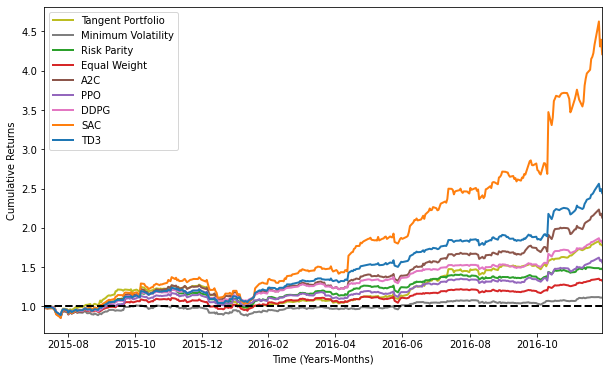}
     \end{subfigure}
     \hfill
     \begin{subfigure}[b]{\textwidth}
         \centering
         \includegraphics[width=\textwidth]{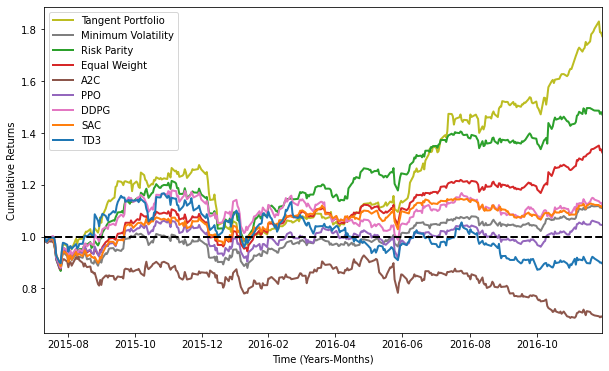}
     \end{subfigure}
        \caption{Evolution of the cumulative returns for the different asset allocation strategies during bull market.
        The results from the DRL-based models are: (Top) The best from 10 runs.
        (Bottom) The worst from 10 runs.}
        \label{fig:bull}
\end{figure}

Figure \ref{fig:bull} plots the cumulative returns of the different asset allocation methodologies proposed in the current work.
While top Figure \ref{fig:bull} shows the run that achieves the best performances, bottom shows the run that obtains the worst.
Furthermore, Table \ref{tab:bull} displays other metrics that help to dissect the methods, to better understand their outcomes.
From all these results, we can derive two main observations: (1) DRL-based models outperform all the traditional approaches, except for PPO, and (2) DRL-based models seem to be unstable, i.e., weight initialization-dependent, and therefore, not as reliable as traditional approaches.
As a result, a hybrid combination between traditional and DRL-based could be a suitable solution since it could find an optimal trade-off: high returns plus stability in the long run.

\begin{table}[!ht]
\centering
\begin{tabular}{lrrrrrrr}

 & \rot{Annual return (\%)} & \rot{Cumulative returns (\%)} & \rot{Annual volatility (\%)} & \rot{Sharpe ratio} & \rot{Calmar ratio} & \rot{Stability} & \rot{Max drawdown (\%)} \\
\hline\hline
Tangent Portfolio &	50.6 & 77.4	& 22.2 & 1.95 & 2.10 & 0.63 & -24.1 \\
Minimum Volatility&	7.7&	11.0&	\textbf{16.4}&	0.53&	0.59	&0.65&	-13.2 \\
Risk Parity	& 31.8&	47.2&	20.8&	1.43&	1.67&	0.84&	-19.1\\
Equal Weight&	22.3&	32.5&	18.1&	1.20&	1.51&	0.82&	-14.7\\
A2C&	71.6&	113.0&	26.5&	2.17&	3.93&	0.92&	-18.2\\
PPO	&37.4&	56.0&	21.7&	1.57&	2.64&	0.86&	-14.2\\
DDPG&	52.8&	81.1&	20.3&	2.19&	4.14&	0.93&	\textbf{-12.7}\\
SAC&	\textbf{179.0}&	\textbf{321.0}&	43.1&	\textbf{2.58}&	\textbf{7.23}&	0.93&	-24.8\\
TD3	&89.2	&144.2&	26.0	&\textbf{2.58}&	5.59& \textbf{0.95}&	-16.0\\
\hline
Tangent Portfolio&	\textbf{50.6}&	\textbf{77.4}&	22.2&	\textbf{1.95}&	\textbf{2.10}&	0.63&	-24.1\\
Minimum Volatility&	7.7&	11.0&	\textbf{16.4}&	0.53&	0.59&	0.65&	-13.2\\
Risk Parity&	31.8&	47.2&	20.8&	1.43&	1.67&	\textbf{0.84}&	-19.1\\
Equal Weight&	22.3&	32.5&	18.1&	1.20&	1.51&	0.82&	-14.7\\
A2C&	-23.3&	-31.1&	23.7&	-1.00&	-0.74&	0.48&	-31.4\\
PPO&	3.2&	4.4&	19.2&	0.26&	0.21&	0.14&	-15.1\\
DDPG&	8.5&	12.1&	21.1&	0.49&	0.52&	0.15&	-16.2\\
SAC&	7.9&	11.2&	17.3&	0.53&	0.69&	0.61&	\textbf{-11.5}\\
TD3&	-7.5&	-10.4&	24.5&	-0.20&	-0.30&	0.47&	-25.3\\
\hline
\end{tabular}
\caption{Financial metrics for the different asset allocation strategies during bull market.
The results from the DRL-based models are: (Top) The best from 10 runs.
(Bottom) The worst from 10 runs.}
\label{tab:bull}
\end{table}

\clearpage

\subsection{Bear Market Scenario}
Bear market occurs when stock prices fall 20\% or more for a sustained period of time.
Triggered by periods of economic slowdown and higher unemployment rate, investors are reluctant to buy, often fleeing for the safety of cash or fixed-income securities.
The result is a seller's market.
Bear markets can last from a few weeks to several years.
The first and most famous bear market was the great depression.
The dot com bubble in 2000 and the housing crisis of 2007–2008 are other examples.

\begin{figure}[h]
  \centering
  \includegraphics[width=\linewidth]{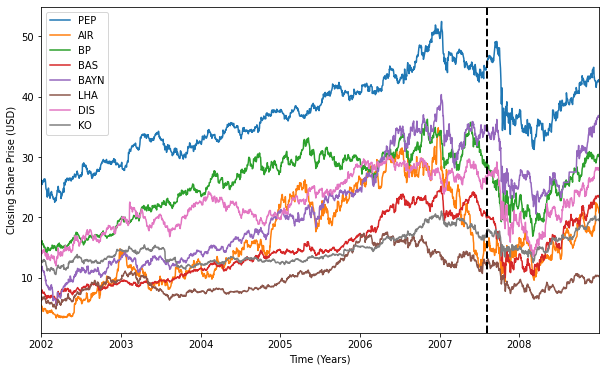}
  \caption{Evolution of the closing share price of our set of assets during bear market; from the 1st of January 2003 to the 1st of January 2010.
  The vertical dashed line separates the training data (left) from the testing data (right).}
  \label{fig:bear_data}
\end{figure}

Our second case of study focuses on 8 stocks that follow a bearish trend.
These securities belong to the following companies:  PepsiCo (PEP), AAR Corp (AIR), British Petroleum (BP), BASF (BAS), Bayer AG (BAYN), Lufthansa (LHA), The Walt Disney Company (DIS) and The Coca-Cola Company (KO).
We use adjusted close prices from the 1st of January 2003 to the 1st of January 2010 (see Figure \ref{fig:bear_data}).

\begin{figure}
     \centering
     \begin{subfigure}[b]{\textwidth}
         \centering
         \includegraphics[width=\textwidth]{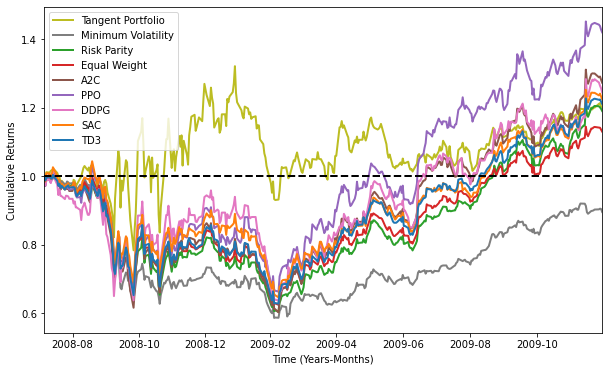}
     \end{subfigure}
     \hfill
     \begin{subfigure}[b]{\textwidth}
         \centering
         \includegraphics[width=\textwidth]{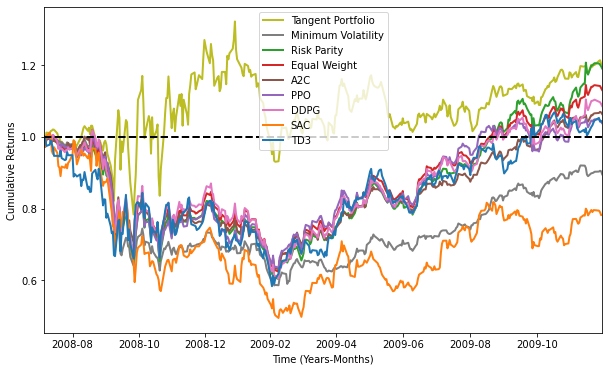}
     \end{subfigure}
        \caption{Evolution of the cumulative returns for the different asset allocation strategies during bear market.
        The results from the DRL-based models are: (Top) The best from 10 runs.
        (Bottom) The worst from 10 runs.}
        \label{fig:bear}
\end{figure}

Figure \ref{fig:bear} plots the cumulative returns of the different asset allocation methodologies proposed in the current work.
While top \ref{fig:bear} shows the run that achieves the best performances, bottom shows the run that obtains the worst.
Furthermore, Table \ref{tab:bear} displays other metrics that help to dissect the methods, to better understand their outcomes.
In Figure \ref{fig:bear}, we can observe that although all methods, except for minimum volatility, end up with positive returns, only tangent portfolio is able to deliver positive returns during (almost) the whole testing period.
Moreover, in this bear setup, the gap between traditional and DRL-based approaches is much reduced, raising concerns of the utility of DRL in declining markets, with the exception of the PPO algorithm.

\begin{table}[!ht]
\centering
\begin{tabular}{lrrrrrrr}

 & \rot{Annual return (\%)} & \rot{Cumulative returns (\%)} & \rot{Annual volatility (\%)} & \rot{Sharpe ratio} & \rot{Calmar ratio} & \rot{Stability} & \rot{Max drawdown (\%)} \\
\hline\hline
Tangent Portfolio&	13.6&	20.2&	51.3&	0.50&	0.46&	0.23&	\textbf{-29.6} \\
Minimum Volatility&	-7.4&	-10.4&	\textbf{27.2}&	-0.14&	-01.8&	0.00&	-41.6 \\
Risk Parity&	12.8&	18.9&	33.1&	0.53&	0.32&	0.27&	-39.6 \\
Equal Weight&	8.9&	13.0&	33.5&	0.42&	0.23&	0.27&	-37.8 \\
A2C&	18.2&	27.3&	42.1&	0.61&	0.45&	0.43&	-40.5 \\
PPO&	\textbf{27.6}&	\textbf{42.0}&	41.4&	\textbf{0.79}&	\textbf{0.75}&	\textbf{0.59}&	-36.8 \\
DDPG&	16.4&	24.5&	52.2&	0.55&	0.44&	0.44&	-37.4 \\
SAC&	15.1&	22.4&	37.1&	0.56&	0.40&	0.32&	-38.1 \\
TD3&	14.2&	21.1&	36.9&	0.54&	0.38&	0.37&	-37.4 \\
\hline
Tangent Portfolio&	\textbf{13.6}&	\textbf{20.2}&	51.3&	0.50&	\textbf{0.46}&	0.23&	\textbf{-29.6} \\
Minimum Volatility&	-7.4&	-10.4&	\textbf{27.2}&	-0.14&	-01.8&	0.00&	-41.6 \\
Risk Parity&	12.8&	18.9&	33.1&	\textbf{0.53}&	0.32&	\textbf{0.27}&	-39.6 \\
Equal Weight&	8.9&	13.0&	33.5&	0.42&	0.23&	\textbf{0.27}&	-37.8 \\
A2C&	4.4&	6.3&	31.2&	0.29&	0.11&	0.16&	-38.2 \\
PPO&	2.6&	3.8&	31.8&	0.24&	0.07&	0.24&	-37.0 \\
DDPG&	5.9&	8.6&	38.2&	0.34&	0.14&	0.20&	-40.6 \\
SAC&	-15.7&	-21.8&	42.4&	-0.19&	-0.31&	0.02&	-51.0 \\
TD3&	2.8&	4.1&	48.2&	0.30&	0.07&	0.26&	-41.6 \\
\hline
\end{tabular}
\caption{Financial metrics for the different asset allocation strategies during bear market.
The results from the DRL-based models are: (Top) The best from 10 runs.
(Bottom) The worst from 10 runs.}
\label{tab:bear}
\end{table}

\clearpage

\subsection{Conclusions}

In this work, we explore the potential of using optimization algorithms for the asset allocation task. To that end, we conduct an extensive benchmark study on 9 different algorithms: tangency portfolio, minimum variance portfolio, risk parity, equal weight, A2C, PPO, DDPG, SAC and TD3. We evaluate their efficacy and reliability on different market conditions, i.e., bullish and bearish tendencies.

Traditional approaches, based on Markowitz portfolio, do not require any fitting optimization process (training) since they do not employ learnable parameters.
In our experiments, these models show stable results, achieving competitive performance on both market scenarios.
Among them, tangency portfolio stands out as this method almost always provides the highest annual and cumulative returns as well as the best Sharpe and Calmar ratios.
Risk parity obtains slightly inferior results, except for a few specific cases, where it outperforms the rest.
As for minimum variance portfolio, it excels at keeping a low annual volatility (fulfilling this extra requirement), but in return the other financial metrics are negatively affected.
Finally, equal weight offers surprisingly decent outcomes, taking into consideration that no optimization is involved.
Although all these traditional proposals can successfully deal with stable market environment, they all are sensitive to outliers and abrupt market's changes.
Therefore, in high volatile markets, traditional approaches are not well suited.
A second drawback arises from their specificity. These algorithms were conceived for specific financial scenarios, and thus, they are rigid tools based solely on asset returns.
In case we wanted to consider other relevant technical indicators such as moving average convergence/divergence, these methods would not be our best candidate.

On the other hand, DRL results are more difficult to interpret. While it is true that deep-learning approaches tend to have runs (random seed settings) that surpass their traditional counterparts, they also provide runs with weaker performance.
For example, PPO and SAC achieve overall the best results in both bullish and bearish markets, respectively, nonetheless, none of them can even beat the equal weight strategy when having poor runs.
The reason for such fluctuations is the training process of these models.
In other words, at training time when the algorithms optimize their agents to learn to make a sequence of decisions through the ``trial and error'' process, there are involved stochastic events.
As a result, independent runs might eventually lead to different optimal or suboptimal solutions.
To cope with this flaw, one could use more data that would indirectly help with convergence and stabilize the training process.
Another complementary solution would be to feed the models with more technical indicators or larger input-data windows, leveraging in this manner the intrinsic flexibility of these neural network architectures.

We expect in the upcoming years to see a lot of exciting new research connecting even more to the fields of finance (asset allocation) and deep reinforcement learning.
Namely, we believe that novel self-attention implementations, such as transformers, could further boost the model's performance.
We would also expect that solutions, trained on synthetic data as a proxy, could help learning better features to lead to more reliable and stable results.
Finally, future work that includes ethics and social aspects, like sustainability, needs to be developed. While it is true that some assets are already following certain ethical codes, to the best of our knowledge, there is no DRL-based approach that incorporates such behaviour into its internal running.

\selectlanguage{english}
\FloatBarrier
\bibliographystyle{plainnat}
\bibliography{biblio.bib}

\clearpage
\section{Appendix}

In this subsection, we visualize the in-depth evolution of the weight allocation for each of the previous scenarios.
This helps us to gain insight of the different running.
Additionally, we plot the statistics (mean and standard deviation) of the DRL-based models for 10 independent runs.
\begin{figure}[h!]
     \centering
     \begin{subfigure}{.91\textwidth}
         \centering
         \includegraphics[width=\textwidth]{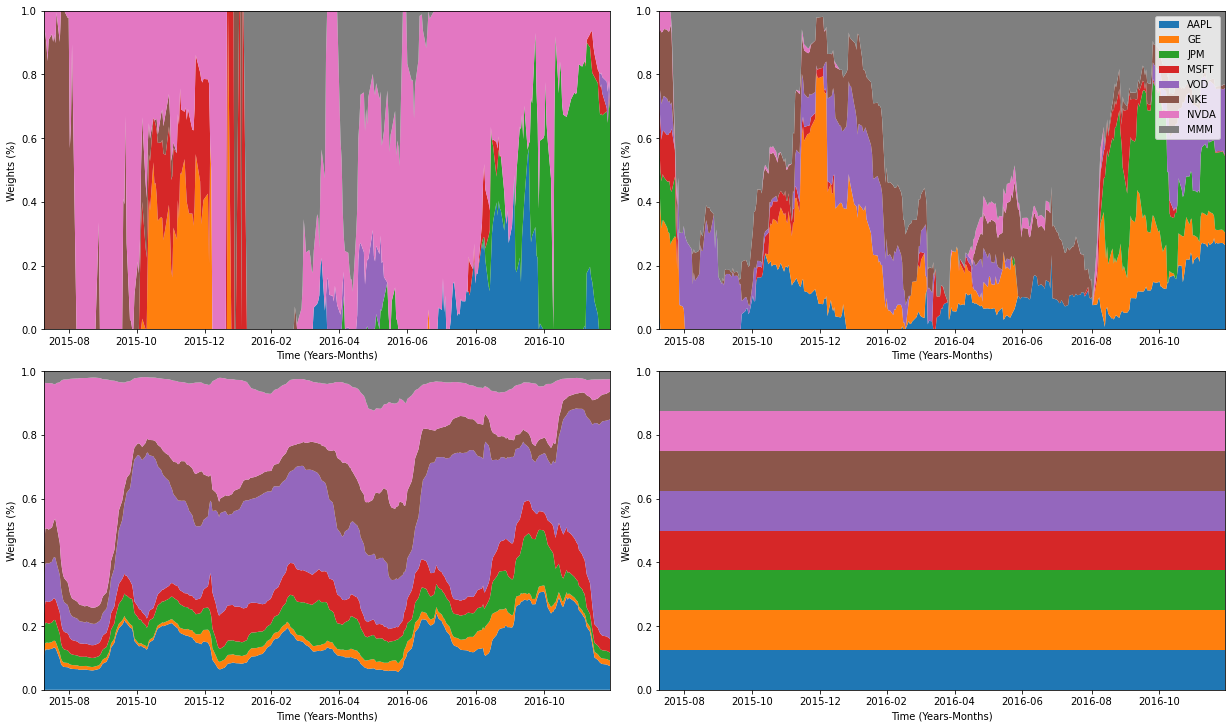}
     \end{subfigure}
     \hfill
     \begin{subfigure}{.91\textwidth}
         \centering
         \includegraphics[width=\textwidth]{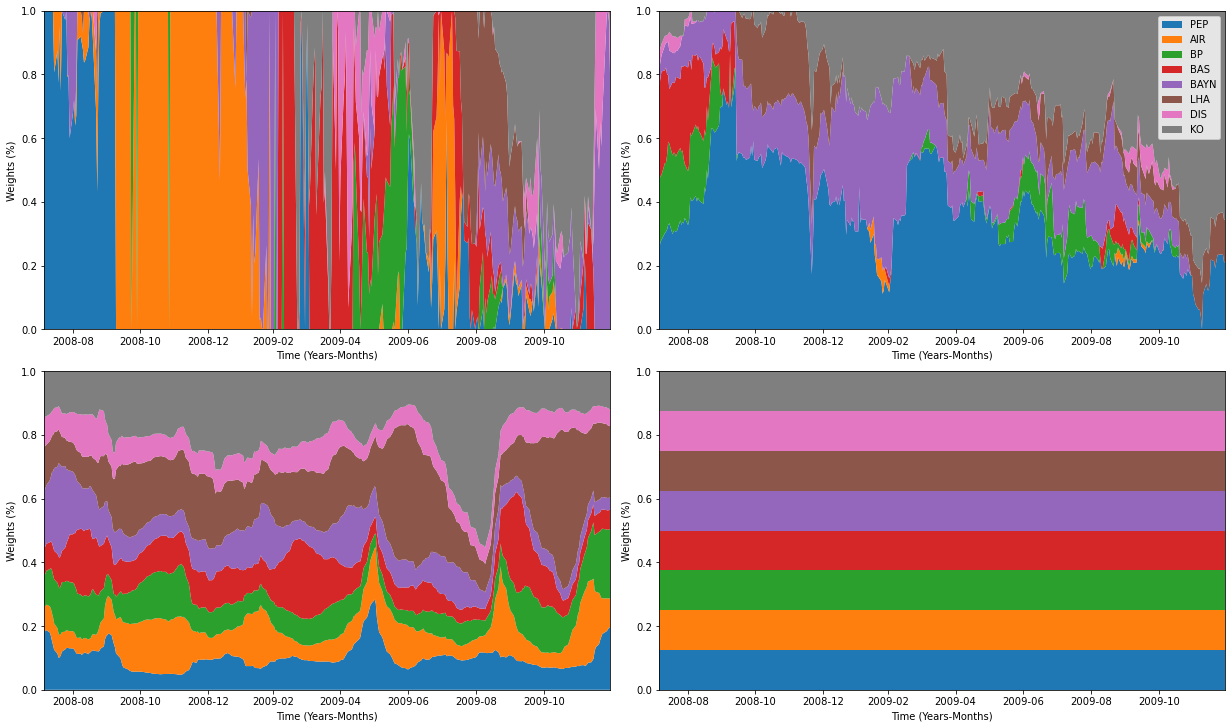}
     \end{subfigure}
        \caption{Weights for assets for the traditional methods.
        Top: bull market.
        Bottom: bear market.
        Clockwise from top left: tangent portfolio, minimum volatility, equal weight, and risk parity.}
        \label{fig:trad}
\end{figure}

\begin{figure}
     \centering
     \begin{subfigure}{.91\textwidth}
         \centering
         \includegraphics[width=\textwidth]{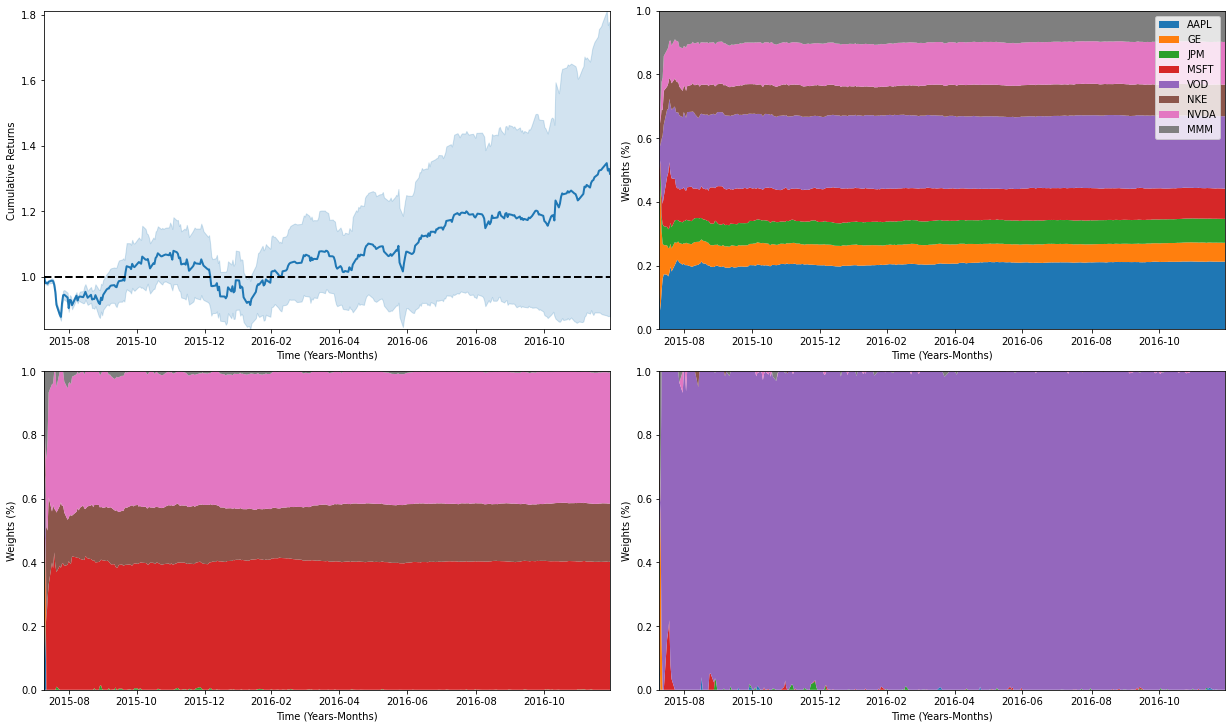}
     \end{subfigure}
     \hfill
     \begin{subfigure}{.91\textwidth}
         \centering
         \includegraphics[width=\textwidth]{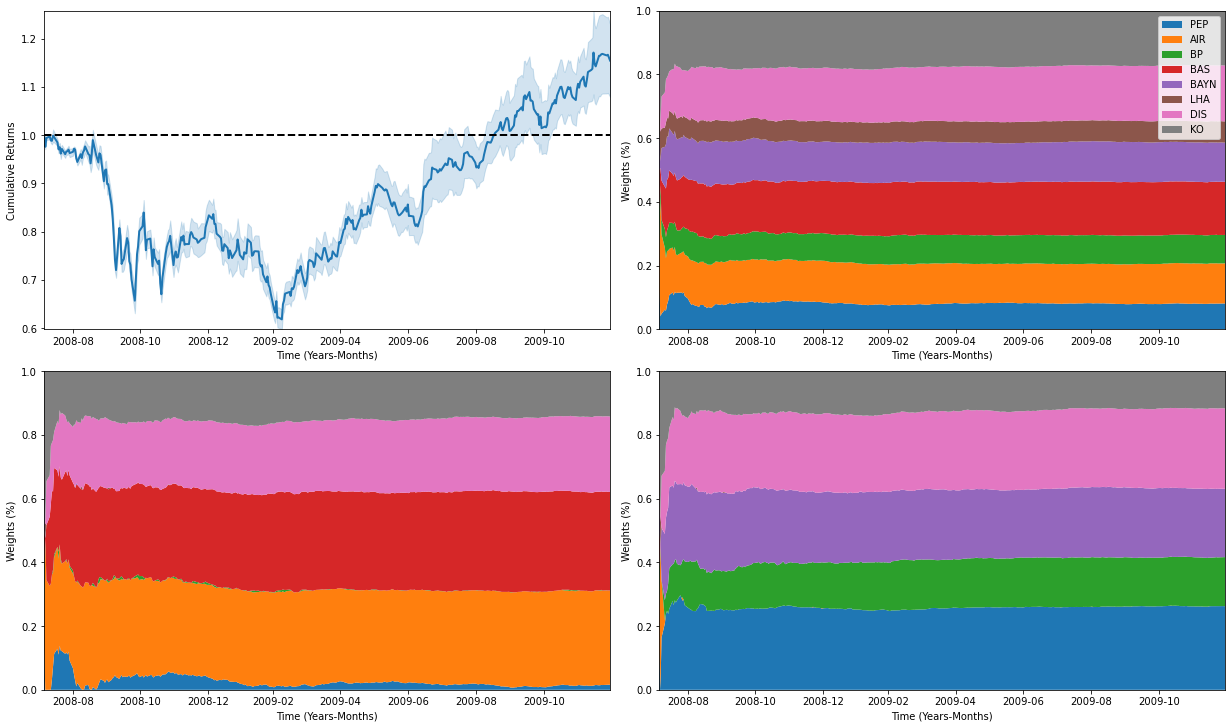}
     \end{subfigure}
        \caption{Evaluation of A2C method.
        Top: bull market.
        Bottom: bear market.
        Clockwise from top left: statistics of the cumulative returns, average weights allocation, worst weight allocation, and best weight allocation.}
        \label{fig:a2c}
\end{figure}

\begin{figure}
     \centering
     \begin{subfigure}{.91\textwidth}
         \centering
         \includegraphics[width=\textwidth]{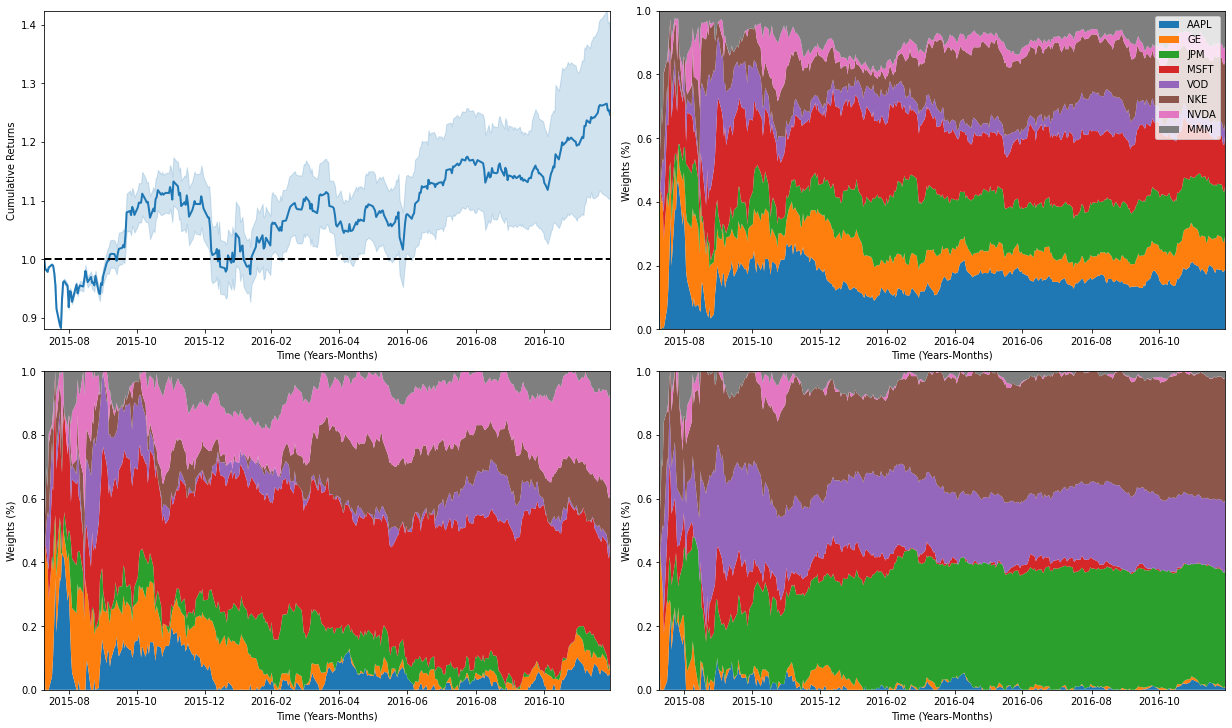}
     \end{subfigure}
     \hfill
     \begin{subfigure}{.91\textwidth}
         \centering
         \includegraphics[width=\textwidth]{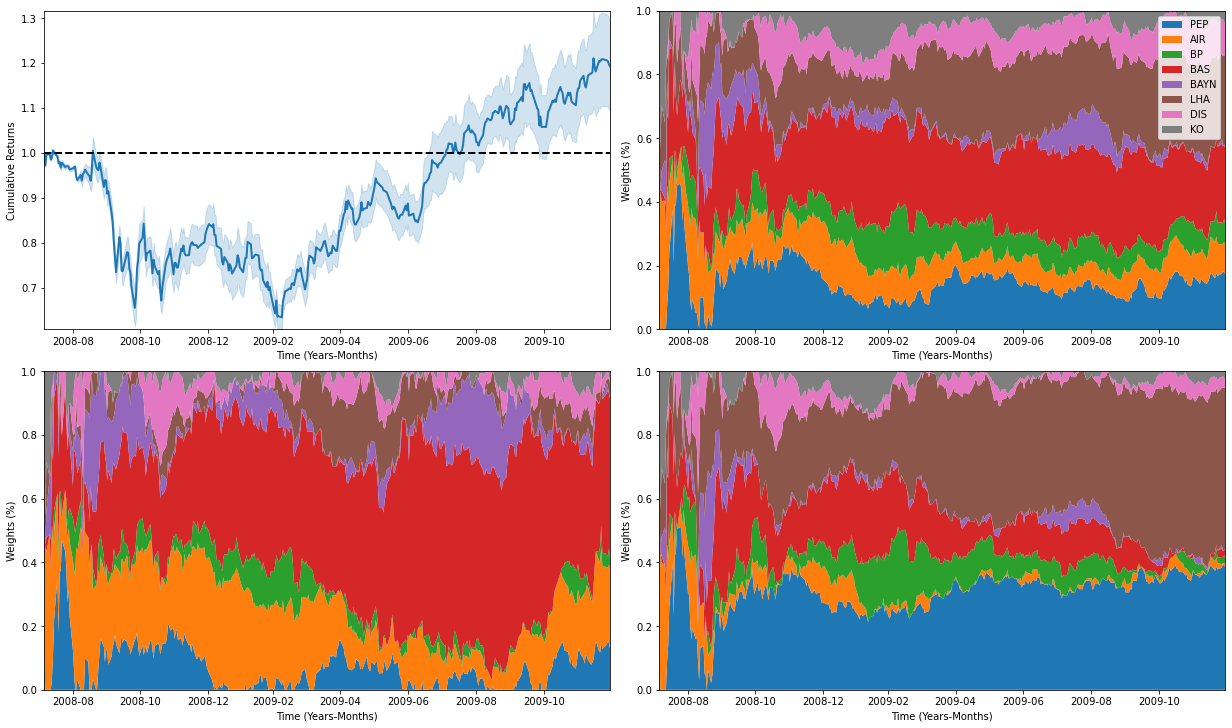}
     \end{subfigure}
        \caption{Evaluation of PPO method.
        Top: bull market.
        Bottom: bear market.
        Clockwise from top left: statistics of the cumulative returns, average weights allocation, worst weight allocation, and best weight allocation.}
        \label{fig:ppo}
\end{figure}

\begin{figure}
     \centering
     \begin{subfigure}{.91\textwidth}
         \centering
         \includegraphics[width=\textwidth]{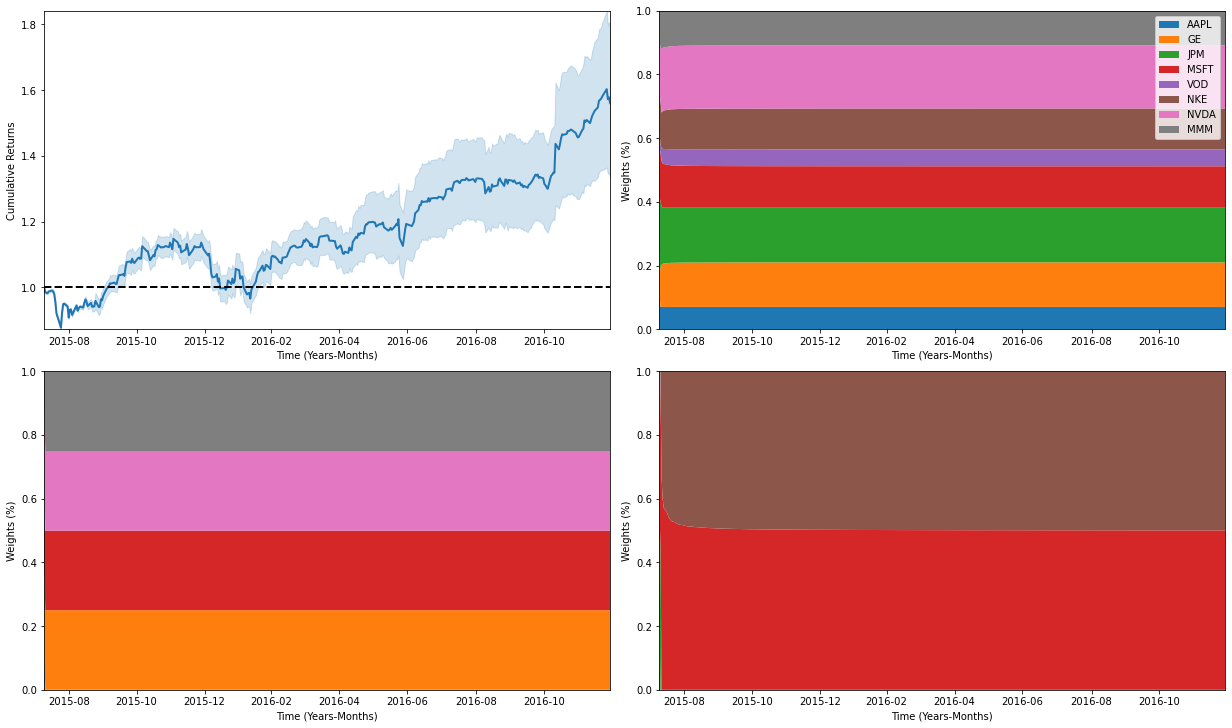}
     \end{subfigure}
     \hfill
     \begin{subfigure}{.91\textwidth}
         \centering
         \includegraphics[width=\textwidth]{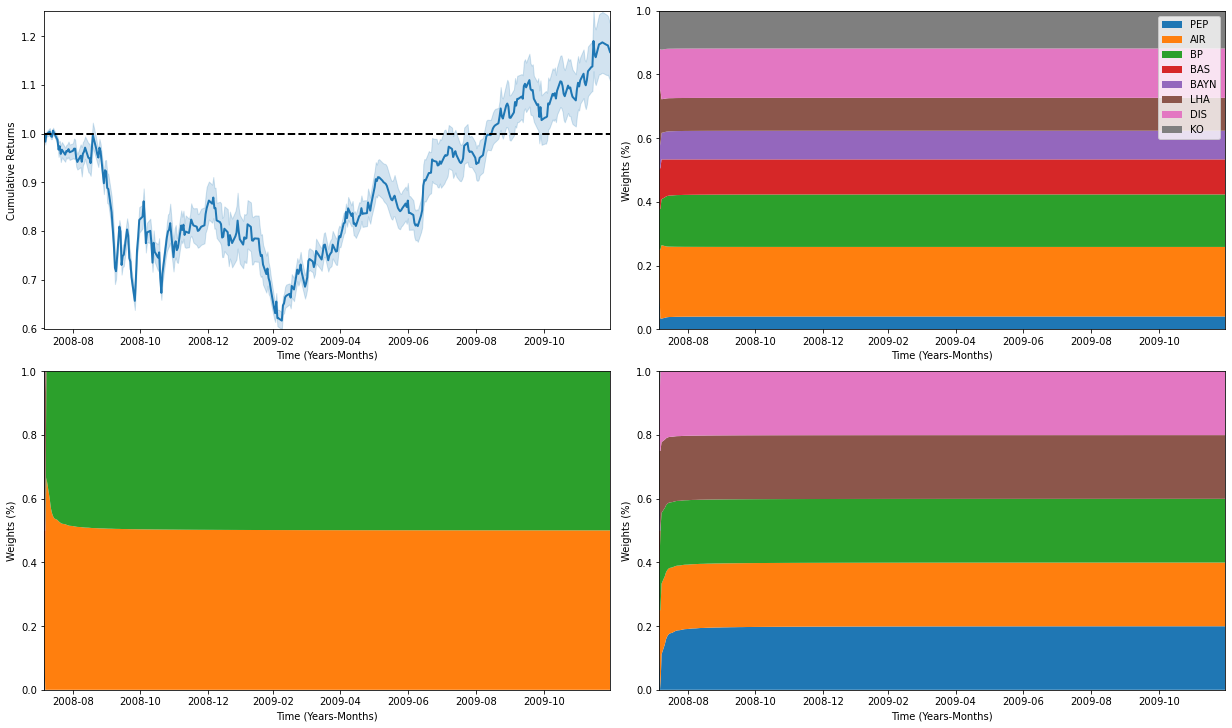}
     \end{subfigure}
        \caption{Evaluation of DDPG method.
        Top: bull market.
        Bottom: bear market.
        Clockwise from top left: statistics of the cumulative returns, average weights allocation, worst weight allocation, and best weight allocation.}
        \label{fig:ddpg}
\end{figure}

\begin{figure}
     \centering
     \begin{subfigure}{.91\textwidth}
         \centering
         \includegraphics[width=\textwidth]{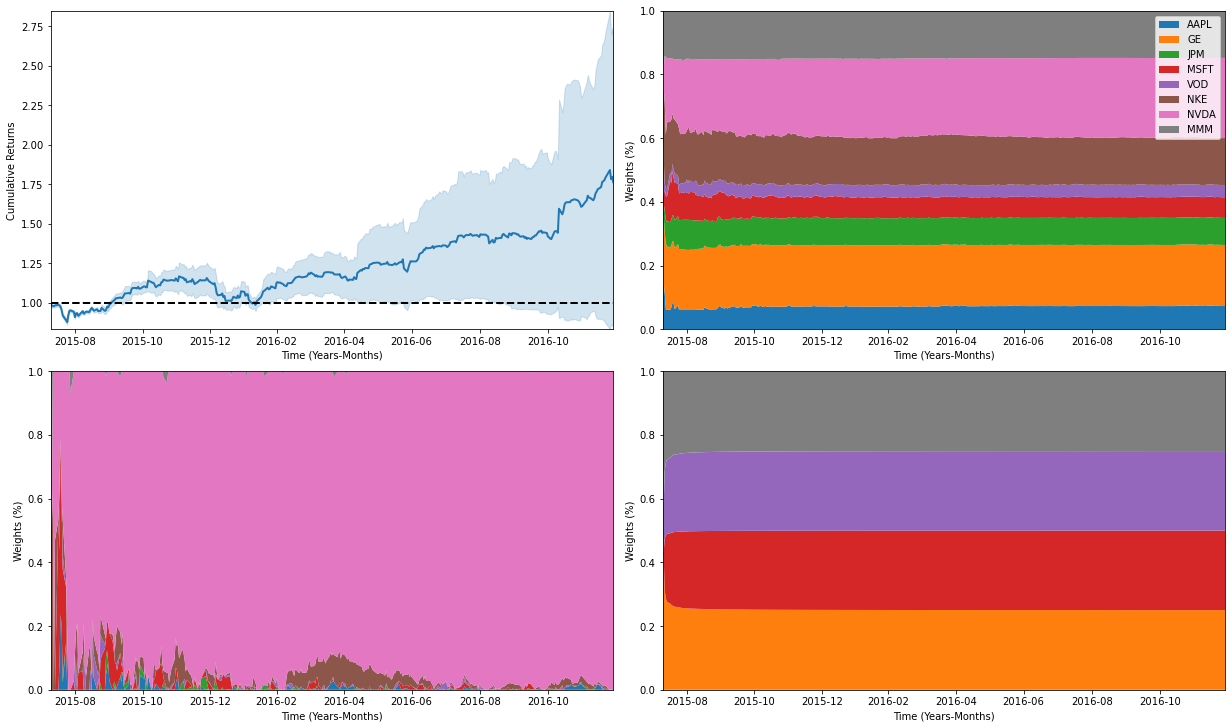}
     \end{subfigure}
     \hfill
     \begin{subfigure}{.91\textwidth}
         \centering
         \includegraphics[width=\textwidth]{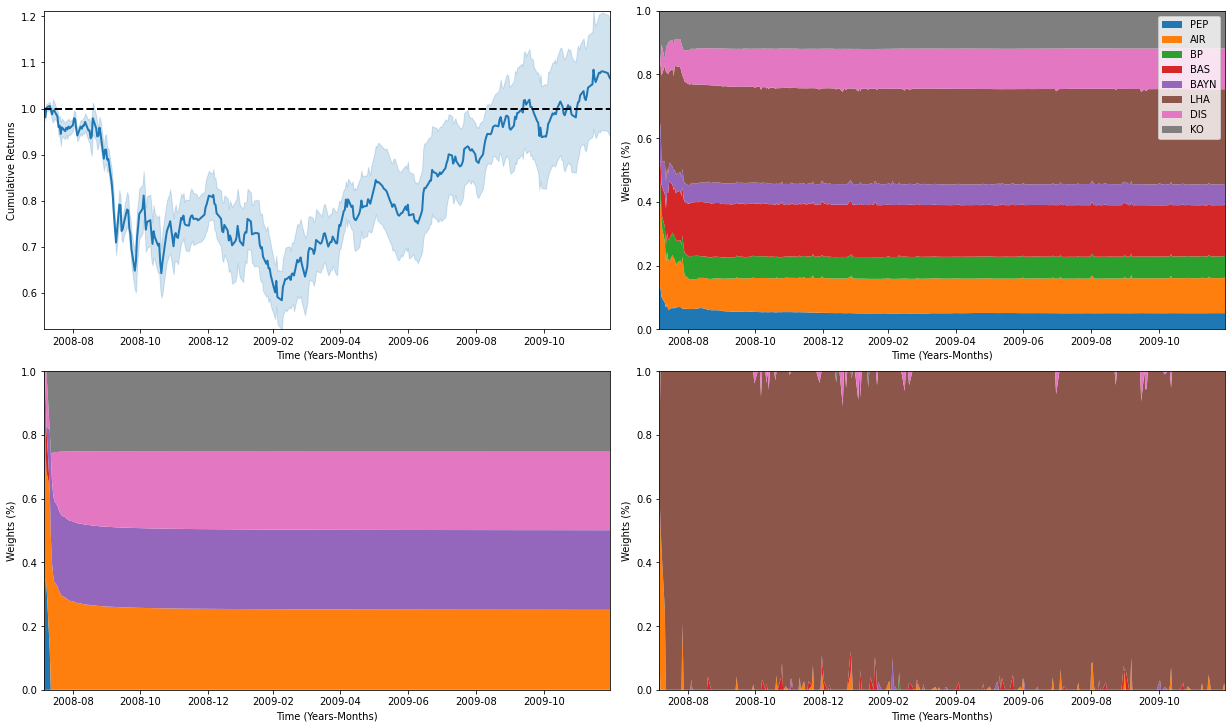}
     \end{subfigure}
        \caption{Evaluation of SAC method.
        Top: bull market.
        Bottom: bear market.
        Clockwise from top left: statistics of the cumulative returns, average weights allocation, worst weight allocation, and best weight allocation.}
        \label{fig:sac}
\end{figure}

\begin{figure}
     \centering
     \begin{subfigure}{.91\textwidth}
         \centering
         \includegraphics[width=\textwidth]{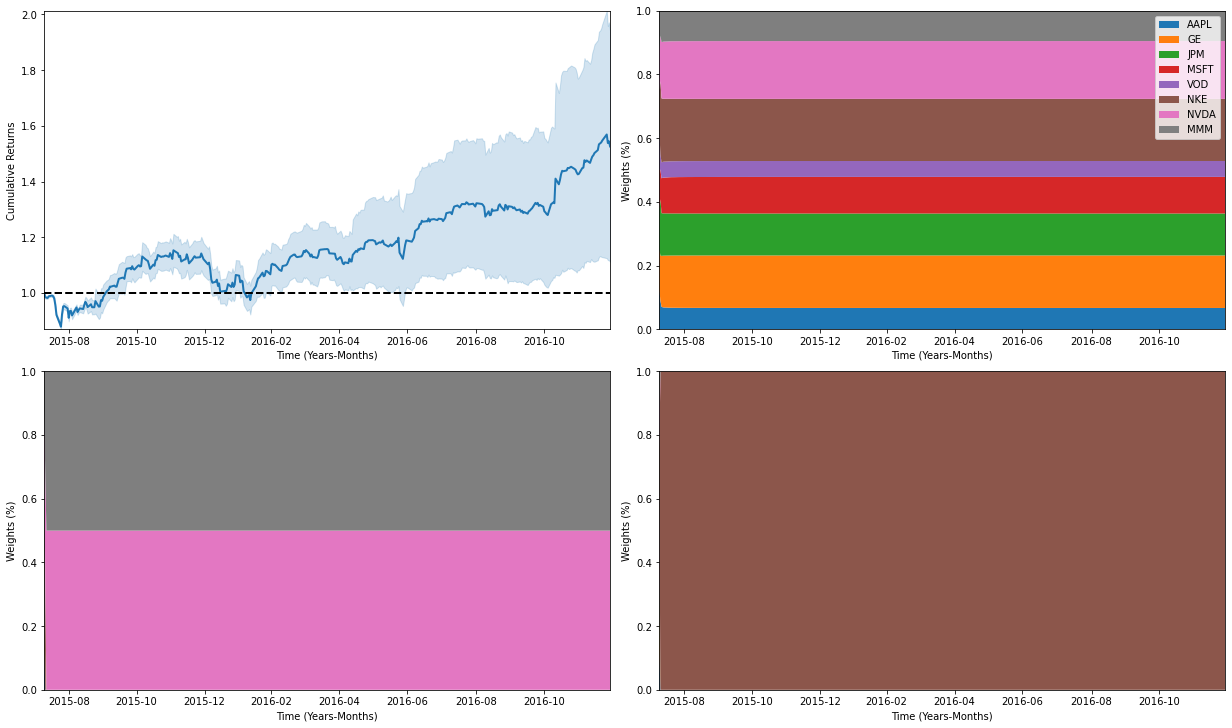}
     \end{subfigure}
     \hfill
     \begin{subfigure}{.91\textwidth}
         \centering
         \includegraphics[width=\textwidth]{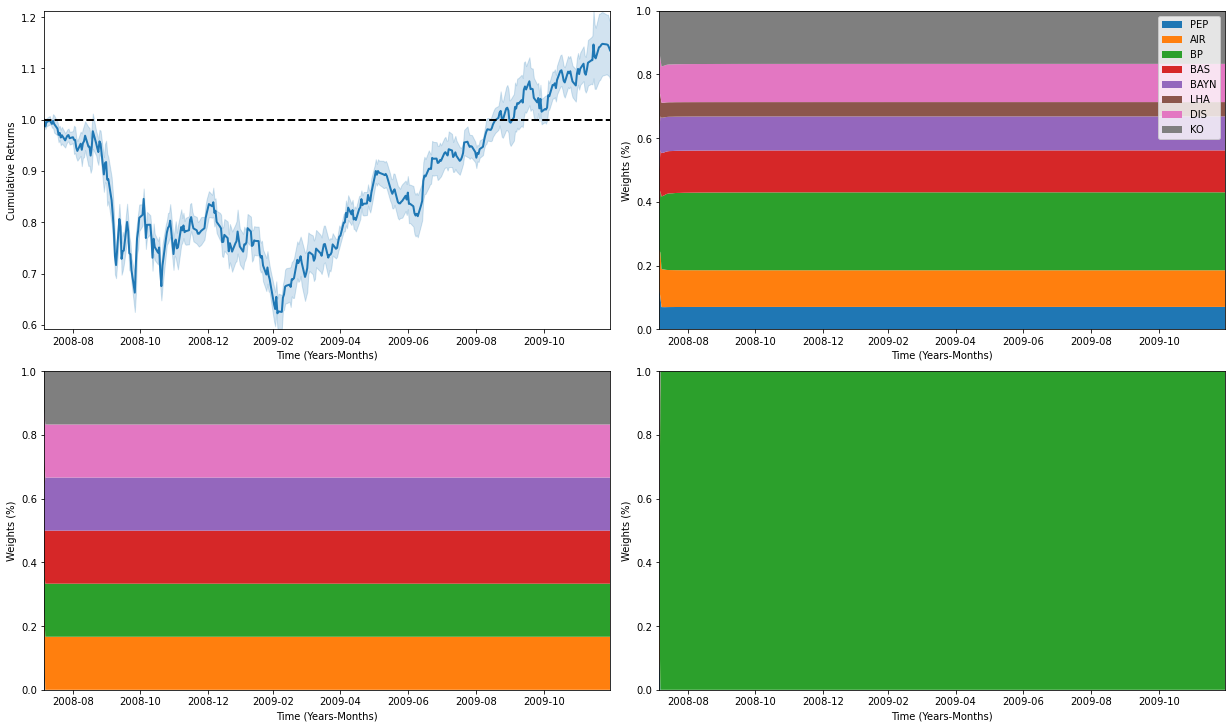}
     \end{subfigure}
        \caption{Evaluation of TD3 method.
        Top: bull market.
        Bottom: bear market.
        Clockwise from top left: statistics of the cumulative returns, average weights allocation, worst weight allocation, and best weight allocation.}
        \label{fig:td3}
\end{figure}

\clearpage
By analysing the evolution of the different weight allocations, we can draw several conclusions.
First, in general, all the DRL-based models avoid dramatic reallocation, changing the portfolio's configuration in a stepwise fashion.
In other words, there are no big differences in the assets' allocation between two consecutive portfolios.
This is, however, an expected behaviour since these models are trained with a transaction cost that penalizes reallocation.
On the other hand, traditional approaches, except for the equal weight, display much dynamic reallocation as no transaction costs are included in their algorithms.
Second, in the bull market, all DRL-based proposals have substantial variations among different runs, which only some of them result into optimal performance.
Nonetheless, we notice that this is not the case for the bear scenario, since the results show a smaller variation, and thus, a more reliable and stable running.
Finally, it is interesting to observe that often the worst runs are associated with ``mono'' asset portfolios (see Figure \ref{fig:a2c}, Figure \ref{fig:sac} and Figure \ref{fig:td3}), i.e., allocation with only one asset.

\end{document}